\documentclass[useAMS,usenatbib]{mn2e}
\usepackage{graphicx}
\usepackage[utf8]{inputenc}
\usepackage{amsmath}

\title[Outflow dynamics of dust-driven wind models]
{Outflow dynamics of dust-driven wind models and
  implications for
cool envelopes of PNe}
\author[Verbena, Schröder, Wachter]{J.~L.~Verbena\thanks{E-mail:
    jluis@astro.ugto.mx (JLV)}, K.-P.~Schröder, and A.~Wachter\\
  Departamento de Astronomía, Universidad de Guanajuato, Apartado
  Postal 144, 36000 Guanajuato, GTO, Mexico}

\begin{document}

\date{Accepted 2011 Month 00. Received 2011 Month 00; in original form
  2011 March 07}

\pagerange{\pageref{firstpage}--\pageref{lastpage}} \pubyear{2010}

\maketitle

\label{firstpage}

\begin{abstract}
The density profiles of cool envelopes of young Planetary Nebulae 
(PNe) are reminiscent of the final AGB outflow history of the central 
star, so far as these have not yet been transformed by the hot wind and 
radiation of the central star. Obviously, the evolution of the mass loss 
rate of that dust-driven, cool wind of the former giant in its 
final AGB stages must have shaped these envelopes to some extent. 
Less clear is the impact of changes in the outflow velocity. Certainly,
larger and fast changes would lead to significant complications in the
reconstruction of the mass-loss history from a cool envelope's density 
profile.

Here, we analyse the outflow velocity $v_{\rm exp}$ in a consistent set 
of over 50 carbon-rich, dust-driven and well ``saturated'' wind models, 
and how it depends on basic stellar parameters. We find a relation of the 
kind of $v_{\rm exp} \propto (L/M)^{0.6}$. By contrast to the vast changes
of the mass-loss rate in the final outflow phase, this relation suggest 
only very modest variations in the wind velocity, even during a thermal 
pulse. Hence, we conclude that the density profiles of cool envelopes around
young PNe should indeed compare relatively well with their recent mass-loss 
history, when diluted plainly by the equation of continuity.
\end{abstract}


\section{Introduction}

The intriguing detail displayed by PNe in visual wavelengths is the result of
complex and dynamic interaction of very different phases of circumstellar
matter, following rapid changes in stellar evolution. The ionizing UV radiation
and hot, fast wind of the hot central star modifies a cool outer envelope from
inside--out \citep{sjspca2005}. In fact, many circular and elliptical PNe often
show a double-shell structure which consists of a rim -- seen as a bright inner
ring -- and the surrounding shell \citep[][and references therein]{prsv2009}.
The latter is material already heated up by the ionizing radiation. The inner
rim, on the other hand, which appears to move with a velocity of
$\sim$40~km~s$^{-1}$, is caused by the interaction of the hot, fast wind of the
central star with the cool, slowly expanding shell. The latter is the product of
a recently ceased, cool, slow ($\sim$15~km~s$^{-1}$), and dense dust-driven
outflow, which accompanied the final asymptotic giant branch (AGB) stages of the
star.

Renzini once coined the nickname ``superwind'' \citep{r1981} for this final,
massive and dust-rich AGB mass loss. It is a result of the very efficient
interaction of the stellar radiation of a cool AGB giant and the newly formed
dust particles in its atmosphere. This superwind removes of the order of one
solar mass in the course of about 30\,000 years from the stellar envelope, prior
to the very end of the AGB phase. This concerns stars with low- and intermediate
initial mass, of about ($\sim$1--8~$M_\odot$), and their mass-loss rates can
reach from $10^{-5}\,M_{\sun}$~yr$^{-1}$ to $10^{-4}\,M_{\sun}$~yr$^{-1}$.

A physical record of such a cool outflow can be found outside the visible
structures of a young PN. By contrast to its spectacular optical counterpart,
the cool envelope is not at all so easy to observe. Only the remarkable advances
of observational means in the past two decades in the infrared (ISO satellite in
the 90ies and Spitzer in the past decade) and millimeter-waves (IRAM, large
advances in detector technology, leading to a promising potential of upcoming
ALMA, see e.g.\ \citep{cox2001} have put such cool envelopes now in reach of
human scrutiny.

Still, details of that outflow phase, and how it gives way to the observed
variety of PNe -- like circular, elliptical, or bipolar -- are not yet fully
understood. It seems clear only that the transition occurs when the stars are
developing from the AGB to PNe, i.e.\ when they are in the short proto-planetary
nebula phase \citep{fmszw2006}. Unfortunately, this phase is rare to observe as
it is short-lived (of the order of $10^3$ years). But the younger a PN is, the
less has the evidence of its recent past -- imprinted in its cool outer envelope
-- been modified by the hot central star.

In addition to all these questions of PN-formation, there is a further
motivation to fully and quantitatively understand the final phase of mass loss
on the AGB: The mass lost during this time determines, by a large proportion,
how much mass actually remains in the stellar remnant, the white dwarf. In other
words, work on superwinds also aims to reproduce the observed initial-final mass
relation \citep{w2000}.

For a quantitative understanding of the radial density profile of a young PN's
cool, outer envelope, the mass-loss history is clearly not the only factor of
importance. Equally well we should know, how fast the outflow was and how much
its velocity was varying. Nevertheless, some cool PN envelopes appear to be
reproduced quite well by means of a proper mass-loss history alone, by making
the simplification that on the star's evolutionary time-scale the outflow
velocity did neither change too much nor too fast. This approach was taken by
\citet{prsv2009}, who compared observed density profiles of PNe derived from
Spitzer images to theoretical predictions of stellar evolution calculations
which included a parameter-dependent mass-loss rate. The latter is based on
models for a C-rich, dust-driven, and dense wind. Despite its simplicity, the
assumption of constant outflow velocity still leads to reasonable agreement
between models and observations. Hence, here we attempt a more detailed
investigation of how much the outflow velocity is actually changing in the
dense, C-rich, and dusty envelope of a red giant, close to the very end of its
AGB stage.

There are, of course, immense limitations to the observational evidence of such
objects. Hence, \citet{htt1994} attempted such an investigation of outflow
velocities based on theoretical arguments. For high mass-loss rates, $\dot M >
10^{-5} \,M_{\sun}$yr$^{-1}$, they found a very modest dependence on stellar
quantities: $v_\text{out} \propto L_\star^{0.3} \delta^{0.5} (\dot M/a)^{0.04}$,
where $L_\star$ is the stellar luminosity, $\delta$ the dust-to-gas ratio, and
$a$ the grain size. The latter appears to have a nearly negligible influence on
the outflow velocity. The same study found, observational evidence would match
well such a small dependence of the outflow velocity on stellar parameters,
i.e.\ on the luminosity.

We here want to take a new and different look at this problem by using the
behaviour of the underlying hydrodynamical wind models -- the same, which were
used for the mass-loss parameterisation in \citet{prsv2009} -- with respect to
the outflow velocity. We are checking the assumption of constant outflow
velocity in our models by checking the average velocities of the dust-driven
wind models against dependence on the basic stellar quantities.

\section{Description of the wind models}

For this investigation we use a set of hydrodynamical wind models that include
formation and growth of dust grains which, by radiation pressure, drive massive
outflows. These are characteristic of the ultimate stages of stellar evolution
on the AGB.

The wind models have been obtained with the computer code developed by
\citet{fgs1992}, \citet{wljhs2000}, and references therein. They result from the
selfconsistent solution of the non-linearly coupled system of equations
describing the hydrodynamical and thermodynamical structure of a spherically
symmetric stellar atmosphere (with a pulsating photosphere as an inner boundary
condition, providing an initial mechanical energy input), its chemical
composition, as well as the nucleation, growth, and evaporation of dust grains,
and the radiation of the central star.

The hydrodynamical wind structure (mass density $\rho$ and outflow velocity $v$)
follows from the equation of continuity and the equation of motion which
includes the radiation pressure on dust grains. The law of energy conservation
and radiative transfer determine the temperature structure.

A carbon-rich chemistry is assumed, where oxygen is completely locked in the CO
molecule. The molecular composition is calculated under the assumption of
chemical equilibrium. The formation, growth, and evaporation of carbon grains is
calculated according to the moment method developed by \citet{gs1988} and
\citet{ggs1990}. \citet{sws2003} give a more detailed summary of the physical
assumptions.

From a set of these models we derived a mass-loss formula for stars with solar
element abundances, which have reached the tip of the AGB \citep{wswas2002}. For
that set of wind models, the velocity amplitude of the piston, used to simulate
stellar pulsation, was set to a value of 5~km~s$^{-1}$. That value was found to
be in accordance with a sample of observed C-star lightcurves. The dependence of
the mass-loss rates on the pulsation period was implicitly accounted for by the
use of an observationally determined period-luminosity relation for Mira stars,
the most appropriate class of objects on the tip of the AGB to compare with the
hydrodynamical models. In this way we were able
to represent the mass-loss rate by a simple relation dependent on the stellar
parameters mass $M$, luminosity $L$, and effective temperature $T_{\rm eff}$,
only. In this work, we follow a similar strategy in oder to find such a simple
relation to represent the average wind velocity.

\begin{table}
  \centering
  \caption{Set of Berlin wind models used in this investigations. The piston
    velocity amplitude is fixed to $\Delta v$~=~5~km~s$^{-1}$, see text for
    details.}
  \label{tab:models_B}
  \begin{tabular}{rrrrr|rrrr}
  \hline
$M$ & $T_\mathrm{eff}$  & $L$  & C/O & $P$ &  $\langle\dot M\rangle$
  & $\langle v_{\text{exp}}\rangle$ \\ [1mm]
[$M_{\odot}$] & [K] & [$L_{\odot}$] & [1] & [d] & [$M_{\odot}$/yr] & [km/s] \\
\hline
1.00 & 2600 & 10000 & 1.80 &  650 & 2.6E-05 & 31.8 \\
0.80 & 3000 & 15000 & 1.50 &  650 & 3.0E-05 & 26.1 \\
0.80 & 3000 & 15000 & 1.50 &  800 & 4.1E-05 & 25.4 \\
0.80 & 3000 & 15000 & 1.50 &  300 & 1.7E-05 & 25.0 \\
0.80 & 3000 &  7500 & 1.80 &  650 & 1.0E-05 & 34.2 \\
0.80 & 3000 &  7500 & 1.80 &  450 & 8.3E-06 & 31.6 \\
0.80 & 2200 & 15000 & 1.30 &  300 & 1.1E-04 & 16.6 \\
0.80 & 2600 &  5000 & 1.30 &  400 & 1.3E-05 &  9.1 \\
0.80 & 2600 &  4000 & 1.30 &  400 & 6.5E-06 &  6.5 \\
0.80 & 3000 &  6000 & 1.30 &  400 & 3.8E-06 & 10.2 \\ 
0.80 & 2600 &  7500 & 1.30 &  450 & 2.5E-05 & 11.3 \\
0.80 & 2600 &  5000 & 1.30 &  300 & 7.4E-06 &  8.6 \\
0.80 & 2500 &  6000 & 1.30 &  400 & 1.6E-05 & 10.0 \\
0.80 & 2800 &  5000 & 1.30 &  400 & 6.1E-06 &  9.1 \\
1.20 & 2600 &  7000 & 1.30 &  400 & 6.1E-06 &  7.9 \\
0.80 & 2600 &  7000 & 1.30 &  450 & 1.9E-05 & 10.8 \\
0.80 & 2600 & 12000 & 1.30 &  800 & 7.0E-05 & 16.2 \\
0.80 & 2600 & 15000 & 1.30 & 1000 & 9.9E-05 & 16.6 \\
0.80 & 2600 & 10000 & 1.30 &  640 & 5.0E-05 & 12.8 \\
1.00 & 2600 & 10000 & 1.30 &  640 & 4.3E-05 & 12.6 \\
1.00 & 2800 & 10000 & 1.30 &  640 & 1.8E-05 & 12.9 \\
1.00 & 2900 & 10000 & 1.30 &  578 & 1.3E-05 & 13.1 \\
1.00 & 2900 & 10000 & 1.25 &  578 & 9.6E-06 & 10.5 \\
1.00 & 2900 & 10000 & 1.25 &  578 & 9.6E-06 & 10.5 \\
1.00 & 2800 &  7000 & 1.30 &  400 & 5.4E-06 &  9.7 \\
1.00 & 2800 &  8000 & 1.30 &  400 & 5.1E-06 & 12.0 \\
1.20 & 2800 & 10000 & 1.50 &  400 & 1.6E-05 & 20.6 \\
1.20 & 2800 & 10000 & 1.80 &  400 & 1.3E-05 & 30.9 \\
0.80 & 2600 &  5000 & 1.30 &  350 & 6.4E-06 &  8.4 \\
0.80 & 2600 &  5000 & 1.30 &  500 & 1.3E-05 &  9.3 \\
0.80 & 2600 &  5000 & 1.30 &  600 & 1.6E-05 &  9.6 \\
0.80 & 2600 &  7500 & 1.30 &  300 & 1.0E-05 & 11.8 \\
0.80 & 2600 &  7500 & 1.30 &  600 & 5.1E-05 & 11.8 \\ 
1.00 & 2400 & 12000 & 1.30 &  600 & 7.7E-05 & 14.7 \\
1.00 & 2400 & 12000 & 1.30 &  300 & 2.6E-05 & 13.6 \\
0.80 & 3000 &  7500 & 1.50 &  400 & 9.0E-06 & 21.6 \\
0.80 & 2400 &  7500 & 1.50 &  104 & 1.4E-05 & 21.4 \\
1.20 & 2800 & 10000 & 1.40 &  400 & 9.8E-06 & 16.9 \\
0.80 & 2550 &  7500 & 1.50 &  104 & 6.0E-06 & 20.9 \\
0.80 & 2600 &  3500 & 1.30 &  400 & 4.9E-06 &  5.4 \\
0.80 & 2600 &  7500 & 1.30 &  800 & 3.6E-05 & 13.1 \\
1.00 & 2400 & 12000 & 1.30 &  500 & 5.9E-05 & 13.4 \\
1.00 & 2400 & 12000 & 1.30 &  800 & 7.9E-05 & 14.4 \\
0.63 & 3000 &  8000 & 1.30 &  820 & 3.0E-05 & 14.3 \\
0.70 & 3000 & 12000 & 1.30 & 1100 & 6.2E-05 & 16.6 \\
0.84 & 3000 & 20000 & 1.30 & 1200 & 7.9E-05 & 19.6 \\
0.94 & 3000 & 25000 & 1.30 & 1300 & 8.8E-05 & 20.7 \\
0.63 & 3500 &  8000 & 1.30 &  820 & 1.4E-05 & 13.6 \\
0.80 & 2700 &  5000 & 1.30 &  300 & 3.9E-06 &  8.4 \\
0.80 & 2700 &  5000 & 1.30 &  350 & 5.6E-06 &  9.4 \\
1.00 & 2800 &  6000 & 1.35 &  400 & 4.6E-06 & 10.4 \\
0.63 & 3500 &  8000 & 1.30 &  460 & 5.8E-06 & 12.7 \\
0.70 & 3500 & 12000 & 1.30 &  650 & 1.0E-05 & 19.5 \\
0.70 & 4700 & 12000 & 1.30 &  650 & 6.0E-06 & 15.7 \\
0.84 & 3500 & 20000 & 1.30 &  880 & 2.0E-05 & 20.6 \\
0.84 & 3700 & 20000 & 1.30 &  710 & 1.1E-05 & 16.1 \\
0.94 & 3500 & 25000 & 1.30 & 1000 & 2.3E-05 & 22.0 \\ 
0.94 & 3700 & 25000 & 1.30 &  810 & 1.2E-05 & 18.1 \\
0.94 & 3900 & 25000 & 1.30 & 1300 & 2.5E-05 & 21.9 \\
\hline
\end{tabular}
\end{table}
The model set used for the current investigation is listed in
Table~\ref{tab:models_B}, given are the input parameters stellar mass $M$,
effective temperature $T_\mathrm{eff}$, luminosity $L$, carbon-to-oxygen ratio
C/O, and piston period $P$. The resulting quantities averaged over typically 20
periods are mass-loss rate $\langle\dot M\rangle$, outflow velocity $\langle
v_{\text{exp}}\rangle$, dust-to-gas ratio $\langle
\frac{\rho_{\text{dust}}}{\rho_{\text{gas}}}\rangle$, and
radiative-to-gravitational acceleration ratio $\langle
\frac{a_{\text{rad}}}{a_{\text{grav}}}\rangle$.

The carbon-to-oxygen ratio of these models is 1.3, which is in good agreement
with both, recent stellar evolution calculations (see \citep{wf2009}, Fig.~1)
and the upper end of the observed C/O range \citep[see for instance][]{bc2005}.
According to both accounts, the C/O ratio rises with the final phases of stellar
evolution, from just above 1 when entering the carbon star phase, to an extreme
of around 1.3 on the very tip of the AGB, where we apply our dust-driven wind
models.

\section{Behaviour of the wind outflow velocity}

\begin{figure*}
  \includegraphics[angle=270,width=.95\textwidth]{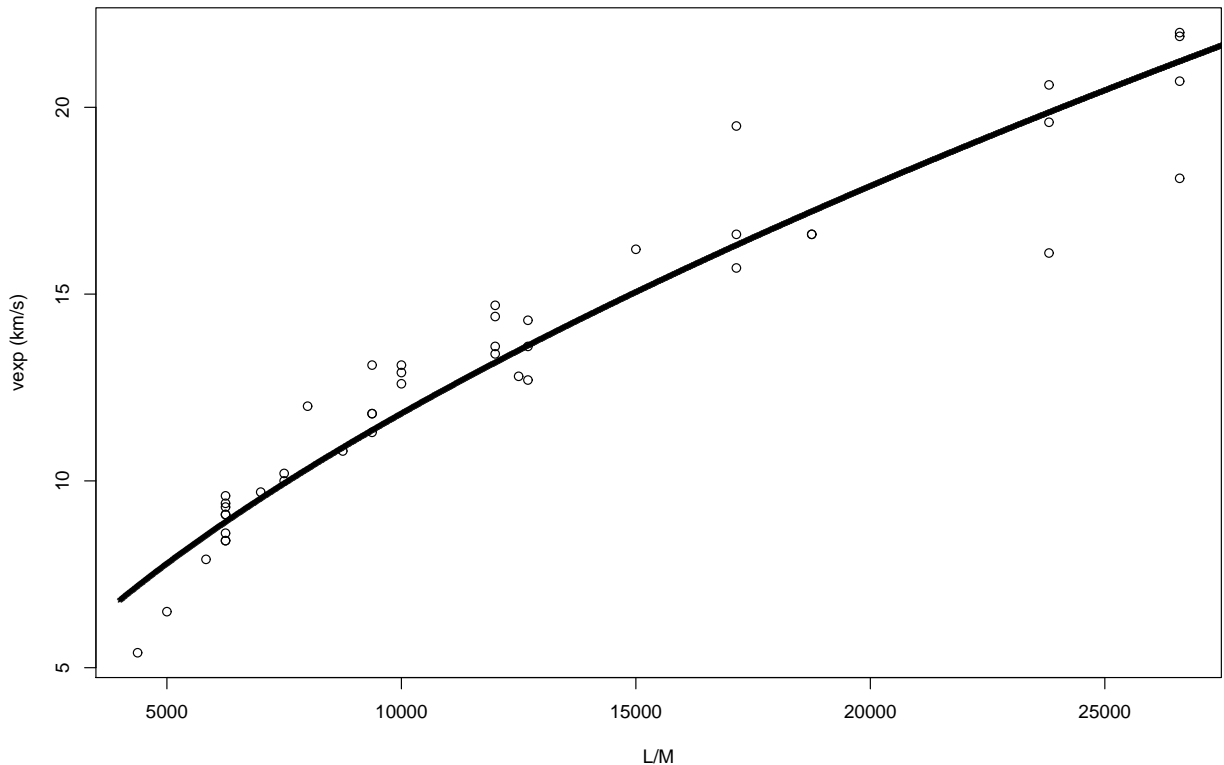}
  \caption{Outflow velocities for our selection of wind models with
    carbon-to-oxygon ratio C/O~=~1.3; the solid line shows
    relation~(\ref{eq:vexp}).}
  \label{fig:velosB}
\end{figure*}

Following simple physical arguments, the dynamics of optically thick winds
should primarily depend on $L/M$, since this ratio is proportional to the
above-mentioned radiative-to-gravitational acceleration ratio $\langle
\frac{a_{\text{rad}}}{a_{\text{grav}}}\rangle$. In fact, \citep{sws1999} find
that truly dust-driven outflows require an Eddington-like lower limit in the
form of a critical $L/M$, above which the mean ratio of radiative to
gravitational acceleration becomes $\langle\alpha\rangle > 1$.

Motivated by this example of the relevance of $L/M$, Fig.~\ref{fig:velosB}
depicts the time-averaged outflow velocities as a function of $L/M$, using the
models detailed in Table~\ref{tab:models_B}, which are restricted to cases of
dust-driven mass loss (characterized by $\langle\alpha\rangle > 1$, see above).
Hence, Fig.~\ref{fig:velosB} supports that there is a relation between the
average wind velocity and the physical parameter of $L/M$.

Based on this approach, we used the non-linear least square fit in the R-base
package \citep{rdct2008}
to derive the dependence of the outflow velocity on $L/M$ in terms of a
power function ($L$ and $M$ in solar units, $v_\text{exp}$ in km~s$^{-1}$):

\begin{equation}\label{eq:vexp}
  v_\text{exp} = 0.05(\pm0.02) \left(\frac{L}{M}\right)^{0.57(\pm0.03)}
\end{equation}

This relation is depicted in Fig.~\ref{fig:velosB} as a solid line. The
relatively small exponent reflects the modest variation of the outflow velocity
over the whole range of wind models depicted in Fig.~\ref{fig:velosB}.

\subsection{Velocity changes induced by stellar evolution}

In the next step, we model the changes of the mean wind velocity, which are
induced by stellar evolution, driven by the related changes in stellar
luminosity $L$ and mass $M$ on the tip of the AGB. In this context, a most
critical case is the immediate aftermath of a thermal pulse (TP), where velocity
changes could be suspected to occur on timescales rapid enough to compete with
the dynamical timescale of the circumstellar envelope (several thousand years).
For the purpose of studying this problem in more detail, we combine the above
velocity-relation (\ref{eq:vexp}) with stellar evolution models, which consider
the respective mass-loss of our dust-driven wind models at each time-step and
yield $L$ and $M$, accordingly.
 
\begin{figure}
  \includegraphics[angle=270,width=.45\textwidth]{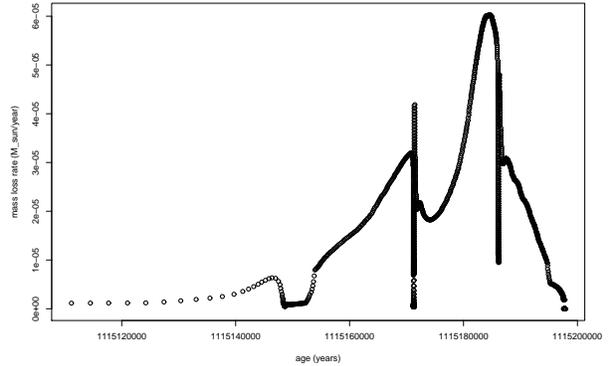}
  \caption{Mass-loss history of a star with solar element abundances of initial
    mass $M_\text{i} = 2.25 \,M_{\sun}$.}
  \label{fig:ML2p25}
\end{figure}

As in earlier work \citep[see][and references therein]{prsv2009} we use the
well-calibrated, fast stellar evolution code originally developed by
\citet{e1973}. The parameterized mass-loss description applied in this code
\citep{wswas2002} is based on the same set of wind models as the current
investigation. As an example, Fig.~\ref{fig:ML2p25} shows the last 80\,000 years
of mass-loss history of a $2.25 \,M_{\sun}$ star with solar element composition
where we can see the superwind phase on the tip of the AGB with a duration of
$\sim$20\,000 years.

\begin{figure}
  \includegraphics[angle=270,width=.45\textwidth]{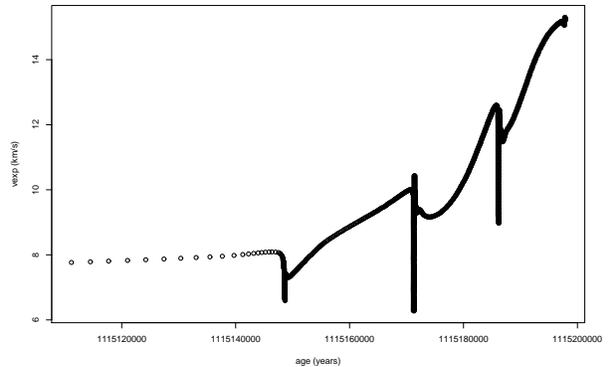}
  \caption{History of outflow velocity according to relation~(\ref{eq:vexp}) of
    same star as in Fig.~\ref{fig:ML2p25}.}
  \label{fig:velo2p25}
\end{figure}

For the same model and timespan, we show the variation of the outflow velocity
in Fig.~\ref{fig:velo2p25}. The mean wind velocity here is seen to grow very
slowly and steadily, with exceptional interruptions caused by thermal pulses. In
these cases, the most prominent change is a temporary decrease, which coincide
with the steep decrease in stellar luminosity, related to the temporary
extinction of the hydrogen burning shell after a helium shell flash. The energy
generated in the latter takes more time to reach the surface. It generates the
slow recovery and further rise of luminosity, mass-loss rate and wind velocity
until the next thermal pulse occurs. In any case, the relative variation of the
wind velocity is very small compared to that of the large simultaneous changes
in the mass loss rate (see Fig.~\ref{fig:ML2p25}). And, important for
hydrodynamic considerations: No large, fast increase of the wind velocity, which
would lead to a compression of the outflow ahead as suggested by colliding wind
scenarios, is seen in Fig.~\ref{fig:velo2p25}.

When considering the superwind phase as a whole, the wind velocity increases
mostly steadily from 8 to 14 km~s$^{-1}$ over a timespan of about 50\,000 years.
This is a lot longer than the dynamic age of any present-day outer cool PN
envelope, which would contain evidence for only the past 10\,000 years or so.
Over that latter order of timespan, only a very modest, systematic increase by
about 1 km~s$^{-1}$ (about 10\%) is expected according to
Fig.~\ref{fig:velo2p25}.

\subsection{The perspective: intrinsic outflow variation}

\begin{figure}
  \includegraphics[width=.45\textwidth]{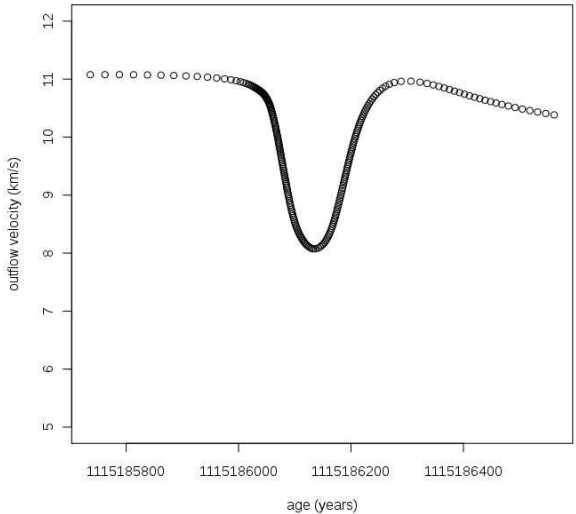}
  \caption{Zoom on last thermal pulse of Fig.~\ref{fig:velo2p25}.}
  \label{fig:velo2p25_TP}
\end{figure}

\begin{figure}
  \includegraphics[width=.45\textwidth]{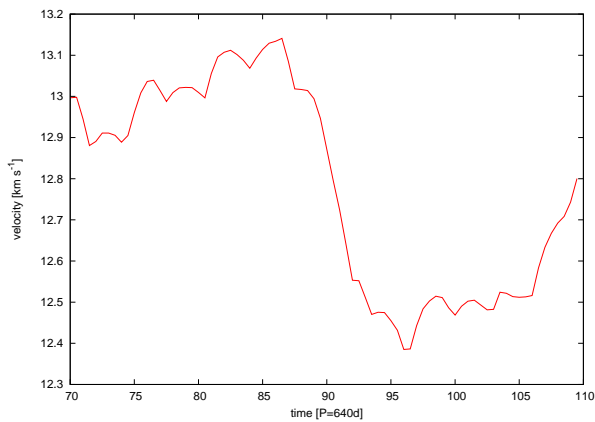}
  \caption{Temporal change over 40 periods of the outflow velocity in a
    hydrodynamical wind model -- input parameters: $M = 1\,M_{\sun}$,
    $T_\text{eff}=2600$~K, $L=10000\,L_{\sun}$, $P=640$~d, C/O=1.3, $\Delta
    v=5$~km~s$^{-1}$.}
  \label{fig:w64_velo}
\end{figure}

As the most rapid and largest velocity change, we identify the about 200 years
after a thermal pulse (TP), when the luminosity goes through a rapid dip. The
outflow velocity corresponds here with a reduction of about 30\%, on a timescale
of only 50 years -- see Fig.~\ref{fig:velo2p25_TP}, which shows the final TP of
Fig.~\ref{fig:velo2p25} over a much smaller timespan.

This is, in expansion timescale, still a modest and, yet, not very fast velocity
change. It does not exceed much the intrinsic, erratic fluctuations of the
outflow of the dynamic wind models, of which Fig.~\ref{fig:w64_velo} shows a
typical example. These frequently reach or exceed 10\%, on a timescale of only
10 years. We should therefore not expect to find any observable evidence for the
TP-related changes of the wind velocity in the density structure of cool PN
envelopes. Rather should the much larger relative changes of the mass-loss rate
be held accountable for any distinguishable radial structure.

\section{Discussion}

\begin{figure}
  \includegraphics[angle=270,width=.49\textwidth]{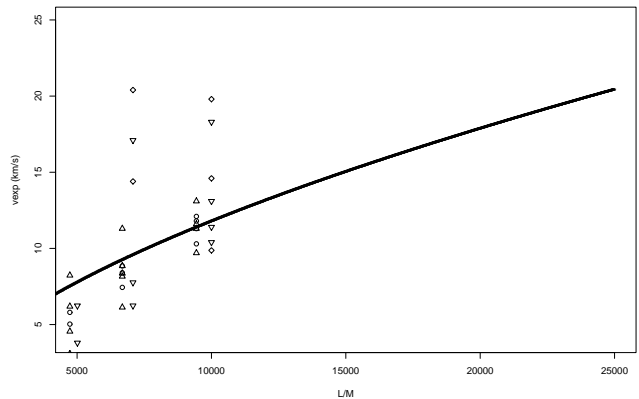}
  \caption{Comparison with wind models by \citet{mwh2010} with C-O~=~8.5
    (corresponds to C/O $\sim$ 1.7); circles and triangles down: $\Delta v =
    4$~km~s$^{-1}$, triangles up and diamonds: $\Delta v = 6$~km~s$^{-1}$, $M =
    0.75$ and $1 \,M_{\sun}$, respectively.}
  \label{fig:velosUpps}
\end{figure}

How do our findings compare with results from other dust-driven wind models?
Recently, \citet{mwh2010} published a grid of wind models with solar element
abundances. The major difference of their models is the inclusion of non-grey
radiative transfer. For comparison, Fig.~\ref{fig:velosUpps} shows the outflow
velocities of those models of \citet{mwh2010}, which have a free carbon
abundance C-O~$= \log[(n_\text{C}-n_\text{O})/n_\text{H}] + 12 = 8.5$ equivalent
to a carbon-to-oxygen ratio of C/O $\sim1.7$. This value is the best option to
compare to our choice of C/O, since only a few models in their grid with C/O
$\sim1.35$, which is closer to our value, result in an outflow. Their $L/M$
ratio does not extend to high values because in their grid models with
luminosities higher than 10\,000~$L_{\sun}$ have only been calculated for masses
larger than 1 $M_{\sun}$.

Furthermore, the models of \citet{mwh2010} show a wider spread in their outflow
velocities -- possibly due the influence of effective temperature. As for the
mechanical energy input, we have set the piston velocity amplitude to
5~km~s$^{-1}$. Hence, for Fig.~\ref{fig:velosUpps} we used the models of
\citet{mwh2010} with the closest choices in this respect, 4 and 6~km~s$^{-1}$.
The largest outflow velocity values seen in the figure correspond to the higher
piston velocity.

Even though their model set seems to suggest a steeper exponent -- as would our
models restricted to the lower L/M range -- our velocity relation overall is
consistent with their grid, especially taking into account the wider spread of
outflow velocity values in the \citet{mwh2010} models.

\section{Conclusions}

From the analysis of a consistent set of dust-driven wind models, we find that
the velocity changes of the superwind can be parameterised by a simple
dependence on $L/M$. In absolute terms, the related velocity changes are of a
small nature, of the order of a fraction of the typical wind velocity. This is
consistent with the very modest dependence of the velocity on $L$ and $M$ found
by \citet{htt1994}, and both these findings are contrasted by the huge changes
of the mass-loss rate (by several orders of magnitude).

Looking at the final phase of the superwind, systematic velocity changes are not
exceeding the inherent short-term fluctuations of the dust-driven wind. And when
a TP falls into this phase, there should only be a 30\% velocity dip on a
timescale of 50 years, which should still be difficult to extract from the
density profile in view of the intrinsic erratic fluctuations of a dust-driven
wind (over 10\% on a timescale of 10 years).

Hence, we may conclude that the huge changes of the mass-loss rate in the
history of a superwind, especially in the aftermath of a TP, are by far the
dominant factor in shaping a present-day cool PN envelope, while changes in the
velocity are hardly exceeding the natural fluctuation of a dust-driven wind.
This find gives some support to a interpretation of cool PN envelopes in terms
of a simple, constant outflow history, where the density profile mostly shows
the progressively (in radial direction, that is with outflow age) diluted
superwind and its mass-loss rate evolution.

\section*{Acknowledgments}
We gratefully acknowledge support by Conacyt project 80804/CB-2007-01 (KPS and
JLV) and by Promep (AW).

\bsp

\label{lastpage}

\end{document}